\documentclass[conference]{IEEEtran}

\usepackage{amsmath,amssymb}

\DeclareMathOperator{\EX}{\mathbb{E}}% expected value

\usepackage[overload]{empheq}
\usepackage{graphicx}
\usepackage[export]{adjustbox}
\usepackage[dvips]{epsfig}
\usepackage{epstopdf}
\usepackage[table]{xcolor}
\usepackage{url}
\usepackage{subfig}
\usepackage{multirow}

\newcommand*\xbar[1]{%
  \hbox{%
    \vbox{%
      \hrule height 0.5pt % The actual bar
      \kern0.5ex%           % Distance between bar and symbol
      \hbox{%
        \kern-0.1em%      % Shortening on the left side
        \ensuremath{#1}%
        \kern-0.1em%      % Shortening on the right side
      }%
    }%
  }%
}

\begin{document}

\title{Analysis and development of a novel algorithm \\ for the in-vehicle hand-usage of a smartphone}

\author{\IEEEauthorblockN{Simone Gelmini\IEEEauthorrefmark{1},
Silvia Strada\IEEEauthorrefmark{1},
Mara Tanelli\IEEEauthorrefmark{1}\IEEEauthorrefmark{3}, 
Sergio Savaresi\IEEEauthorrefmark{1}, and
Vincenzo Biase\IEEEauthorrefmark{2}}
\IEEEauthorblockA{\IEEEauthorrefmark{1}Dipartimento di Elettronica, Informazione e Bioingegneria\\
Politecnico di Milano\\
Piazza Leonardo da Vinci 32, 20133 Milan, Italy\\
Email: \{simone.gelmini, silvia.strada, mara.tanelli, sergio.savaresi\}@polimi.it}
\IEEEauthorblockA{\IEEEauthorrefmark{2}Kubris S.r.l., Via Benigno Crespi 57, 20159 MILANO\\
Email: vincenzo.biase@kubris.com}
\IEEEauthorblockA{\IEEEauthorrefmark{3}corresponding author}}

\IEEEspecialpapernotice{(Special session paper - Session ID:H22)}

\maketitle

\IEEEpeerreviewmaketitle

%%%%%%%%%%%%%%%%%%%%%%%%%%%%%%%%%%%%%%%%%%%%%%%%%%%%%%%%%%%%%%%%%%%%%%%%%%%%%%%%
%%%%%%%%%%%%%%%%%%%%%%%%%%%% Abstract %%%%%%%%%%%%%%%%%%%%%%%%%%%%%%%%%%%%%%%%%%
\begin{abstract}
Smartphone usage while driving is unanimously considered to be a really dangerous habit due to strong correlation with road accidents. In this paper, the problem of detecting whether the driver is using the phone during a trip is addressed. To do this, high-frequency data from the tri-axial inertial measurement unit (IMU) integrated in almost all modern phone is processed without relying on external inputs so as to provide a self-contained approach. By resorting to a frequency-domain analysis, it is possible to extract from the raw signals the useful information needed to detect when the driver is using the phone, without being affected by the effects that vehicle motion has on the same signals. The selected features are used to train a Support Vector Machine (SVM) algorithm. The performance of the proposed approach are analyzed and tested on experimental data collected during mixed naturalistic driving scenarios, proving  the effectiveness of the proposed approach.
\end{abstract}

%%%%%%%%%%%%%%%%%%%%%%%%%%%%%%%%%%%%%%%%%%%%%%%%%%%%%%%%%%%%%%%%%%%%%%%%%%%%%%%%
%%%%%%%%%%%%%%%%%%%%%%%%%% Introduction %%%%%%%%%%%%%%%%%%%%%%%%%%%%%%%%%%%%%%%%
\section{Introduction}
Distracted driving due to the use of mobile devices contributes to a significant amount of fatalities per year \cite{ncsa2010report}, attracting the attention not only from government regulators but also from mobile communication and insurance companies. While some network operators and smartphone makers have started to adopt masked approaches that actively seek to manage distraction (\textit{e.g.}, the \textit{Do not disturb while driving} mode introduced by Apple \cite{Apple}), these methods do not prevent the complete use of the phone. Discounts from the insurance companies could be really persuasive. To this end, it is crucial to monitor whether and how often a phone is used when actively driving, better understanding the risk associated with this hazardous habit and the consequent incentives.

In the scientific literature, several works on profiling drivers' habits for safety-oriented purposes have been presented. To this purpose, relevant profiling indexes are extracted by means of the smartphones' intrinsic sensing capability, generally composed of low-cost sensors such as accelerometers, gyroscopes, magnetometers, \textit{etc.}. Commonly employed drivers’ risk profile indexes are based on measures related to speeding, driving smoothness, harsh accelerations, brakes, swerves, and cornerning, \cite{7891981, wahlstrom2015driving,johnson2011driving,eren2012estimating}. In \cite{chen2015d}, it is shown that all most dangerous driving behaviors share a unique pattern in terms of acceleration and orientation. The recognized patterns are used in an abnormal driving behavior detection systems, which perform a real-time abnormal driving behavior monitoring. As proposed in \cite{dai2010mobile}, specific patterns can be mined also when the driver is driving under influence, another particularly risky driving condition. 

However, in the existing literature, little or no effort has been done to detect the use of the phone while driving, which can be for texting, making a phone call or simply accessing the Internet. This problem has been tackled only in few contributions. In \cite{li2016dangerous}, the authors implement a driving-style profiling method which takes into account not only dangerous maneuvers but also phone usage during the trip. The recognition of the latter is accomplished in a quite na\"ive fashion, \textit{i.e.}, by simply looking if significant acceleration or angular velocity changes are sensed by the phone sensors within a short time-window. This enables checking if a pick-up or drop-off of the phone occurred, but not if the device has been really used in between. In \cite{bo2017detecting}, integrated inertial sensors are employed, detecting the simultaneous occurrence of driving and texting in real-time, by mining a well defined pattern. The same problem has been analyzed also by \cite{wang2013sensing,chu2014smartphone,yang2011detecting}, where the detection of the phone is obtained by the fusion of the smartphone built-in sensors and external devices.

In this paper, we explore a novel self-contained approach that dynamically detects when the smartphone is used elaborating only the smartphone inertial sensors measurements, without the need of analyzing device's stats (e.g. the use of CPU and RAM) or smartphone operation system's interrupts (e.g., event listeners). An activating beacon switches on the algorithm when entering the vehicle, thus limiting the monitoring to the journey only, making the proposed algorithm different from other contributions (e.g., \cite{chu2014smartphone}) in which the phone constantly monitors the user. The presented algorithm consists of a binary classification task on time series measured by the smartphone sensors with high sampling rates, which represent three-axis accelerations and three-axis angular velocities. Features are extracted by means of a data-preprocessing phase carried out in the frequency domain, making the algorithm robust with respect to different scenarios and driving conditions. The classification is performed resorting the well-known machine learning (ML) algorithm SVM. To our best knowledge, this is the first work in which high-frequency, real-time signal processing is used for smartphone usage mode detection \textit{via} classification. Despite the simple and cost-effective sensor layout, corrupted by the unavoidable vehicle dynamics disturbances, the proposed approach yields a classification accuracy of $96\ \%$, a sensitivity of $99\ \%$ and a specificity of approximately $90\ \%$.

The paper is organized as follows. Section \ref{set_up} defines the problem and the experimental setup. In Section \ref{features_selection}, data are analyzed, both in time and frequency-domain, leading to design the data preprocessing phase. Five meaningful features are extracted, analyzed and discussed in Section \ref{features_selection}. Section \ref{algorithm}
discusses the classification method. The features selection and classification results are also discussed in Section \ref{results}, showing the effectiveness of the proposed solution. 

%%%%%%%%%%%%%%%%%%%%%%%%%%%%%%%%%%%%%%%%%%%%%%%%%%%%%%%%%%%%%%%%%%%%%%%%%%%%%%%%
%%%%%%%%%%%%%%%% Problem statement and experimental setup %%%%%%%%%%%%%%%%%%%%%%
\section{Problem statement and experimental setup} \label{set_up}
In this paper, a smartphone usage detection algorithm is proposed. The detection is limited to the situation in which the user is inside the vehicle, assumed to be driving. For this purpose, the algorithm is activated (and deactivated) when the driver reaches (or leaves) the driving seat, thanks to a Bluetooth beacon placed under the steering wheel. The proposed algorithm aims at detecting the use of the phone based only on the sensed motion, discriminating between the effects due to pure vehicle dynamics and those related to the phone use.

\begin{figure}[thpb]
 \centering
 \includegraphics[width=0.45\textwidth]{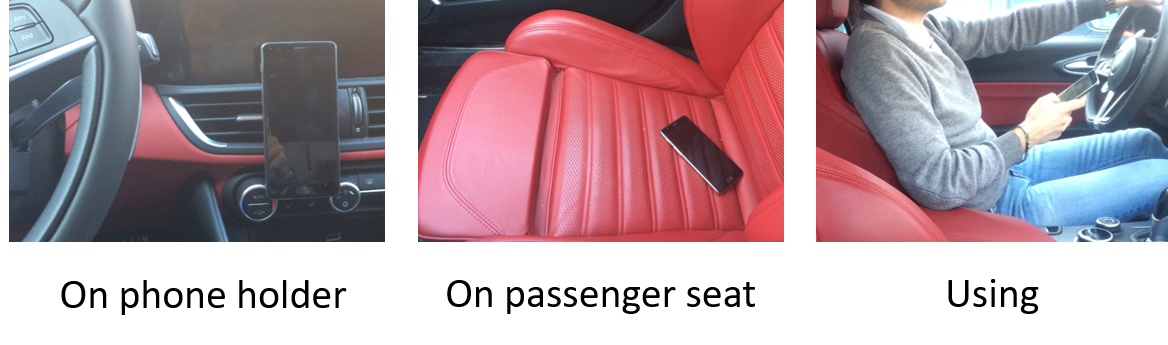}
 \caption{An overview of the real use case scenarios reproduced to train the algorithm.}
 \label{fig:case_studies}
\end{figure}

The smartphone is considered \textit{in-use} when the driver is performing one of the most common activities (\textit{e.g.}, handling, texting, scrolling, browsing, calling, \textit{etc}.) and \textit{not in-use} otherwise. As it would be impossible to fully cover all the possible ways in which the phone can be used, the detection algorithm is trained with a limited number of the most common daily scenarios. For the case of \textit{not in-use}, the considered scenarios include situations in which the phone is on the phone holder or on the passenger seat. On the contrary, in the \textit{in-use} case, all the possible uses of the phone are considered as a single global condition. A list of the considered scenario is graphically described in Fig. \ref{fig:case_studies}.

\begin{figure}[thpb]
 \centering
 \includegraphics[width=0.45\textwidth]{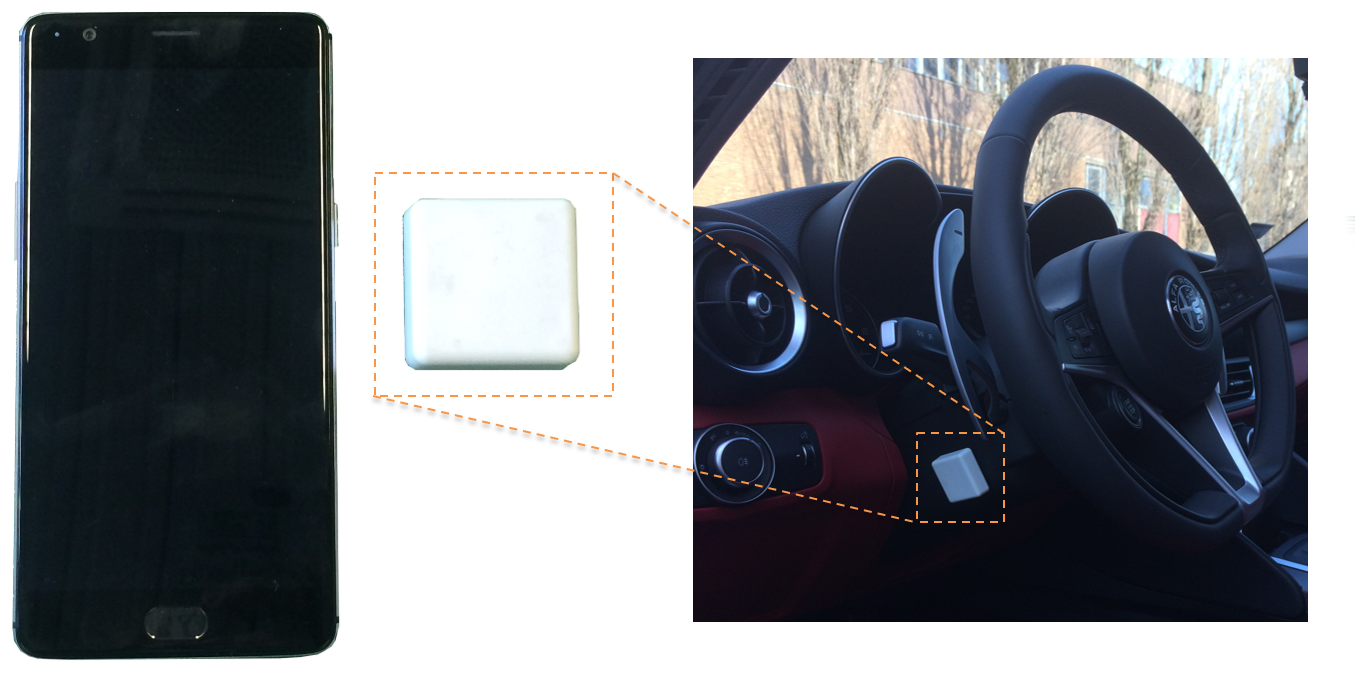}
 \caption{The experimental setup: a OnePlus One smartphone and the Bluetooth beacon that is placed under the steering wheel.}
 \label{fig:experimental setup}
\end{figure}

As shown in Fig. \ref{fig:experimental setup}, the experimental setup is composed of a smartphone only - in which an app records data from the available sensors sensors - and of the activating beacon. It is widely known that smartphones are equipped with many sensors, but in this application only the STMicroelectronics LIS3DH, a tri-axial IMU, is used. Data are recorded at the maximum sampling frequency of $120\ \mathrm{Hz}$. %The classification algorithm, in order to detect the frequency components related to the use of the phone, is fed with data at the maximum sampling frequency, performing a prediction every $0.0083$ seconds. However, since the final goal of detecting the use of the phone is for profiling purposes, the classifier's prediction is averaged over a one second time window, which is consistent with that of common apps performing driving-style assessment in the insurance telematics context.
%%%%%%%%%%%%%%%%%%%%%%%%%%%%%%%%%%%%%%%%%%%%%%%%%%%%%%%%%%%%%%%%%%%%%%%%%%%%%%%%
%%%%%%%%%%%%%%%%%%%%%% Case studies data analysis %%%%%%%%%%%%%%%%%%%%%%%%%%%%%%
\section{Features selection} \label{features_selection}
As described in the previous section, the aim of the algorithm is to detect the general condition of the smartphone being used (irrespectively of the specific action being performed by the user with it). In this section, the necessary steps to extract and select the most informative features from the on-board IMU are presented.

\subsection{Preliminary data analysis}
To extract the most expressive features, a preliminary analysis of the experimental data is needed. As mentioned in Section \ref{set_up}, the data for this application consist of three acceleration and angular velocity measurements. Analyzing six signals separately might be unnecessarily complicated and, in order to be as general as possible, data from different smartphone orientations would be necessary. For these reasons, the information is condensed by computing acceleration and angular speed Euclidean norms as
\begin{equation}
\begin{split}
\|a\|=&\sqrt{a_x^2+a_y^2+a_z^2} \\
\|\omega\|=&\sqrt{\omega_x^2+\omega_y^2+\omega_z^2}
\end{split}
\label{acc_norm}
\end{equation}
thus representing, with two signals only, the overall intensity of both accelerations and angular rates.

\begin{figure}[thpb]
 \centering
 \includegraphics[width=0.45\textwidth]{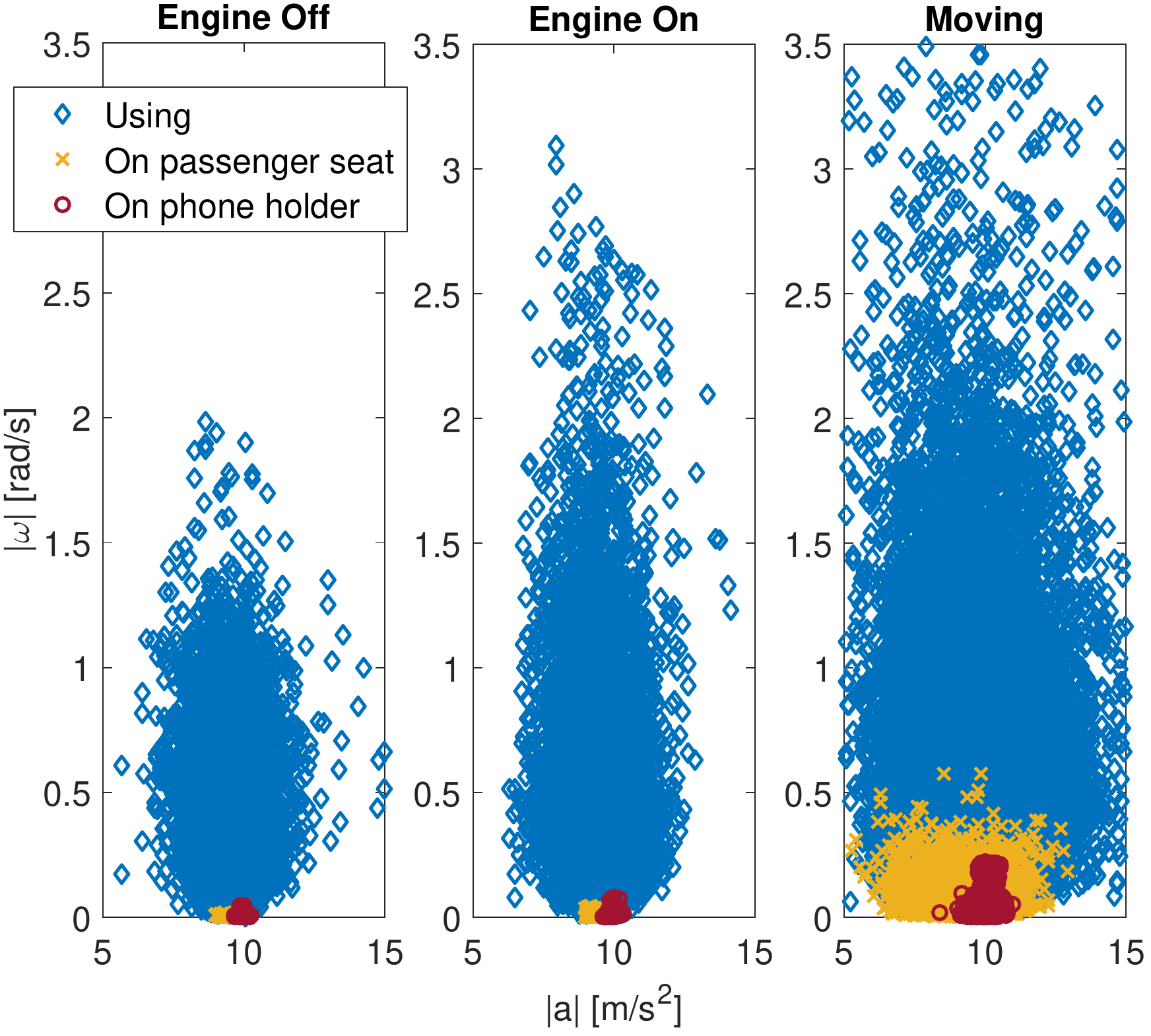}
 \caption{The scatter plots for three scenarios (\textit{using}, \textit{on passenger seat}, \textit{on phone support}) analyzed in three driving situations: when the vehicle is not moving and the engine is on/off, a baseline from the vibration intensity perspective, and when the vehicle is moving. The analysis of the acceleration and angular rate norms is attitude independent, though the motion is described only by two inclusive signals.}
 \label{fig:scatter_plot}
\end{figure}

In Fig. \ref{fig:scatter_plot}, the acceleration and angular velocity norms are shown for different scenarios and for different vehicle conditions: \textit{engine off}, \textit{i.e.}, when the vehicle is powered off; \textit{engine on}, \textit{i.e.}, when the vehicle is powered on, but standing still; \textit{moving}, \textit{i.e.}, when vehicle is in motion. In the \textit{engine off} situations, the acceleration norm has approximately the value of gravity and rotational velocity is negligible, though in \textit{using}, both the acceleration and the angular velocity norms are more excited and span a wider range. In the \textit{engine on} condition, the same situation is experienced, with few samples on the angular velocity norm exceeding $2\ \mathrm{rad/s}$ in case of \textit{using}. The only differences between \textit{on phone support} and \textit{on passenger seat} are due to small sensor drift and noise only. It is possible to say that these two vehicle states are, from the sensed motion perspective, the same for both the considered scenarios.

A different situation is experienced when the vehicle is \textit{moving}. In this case, the vehicle dynamics emphasizes the three clouds in both acceleration and angular rate axes. The \textit{using} scenario is more varied than the others, even if the overlap is consequently greater too, in particular on the acceleration norm axis.

\begin{figure}[thpb]
 \centering
 \includegraphics[width=0.45\textwidth]{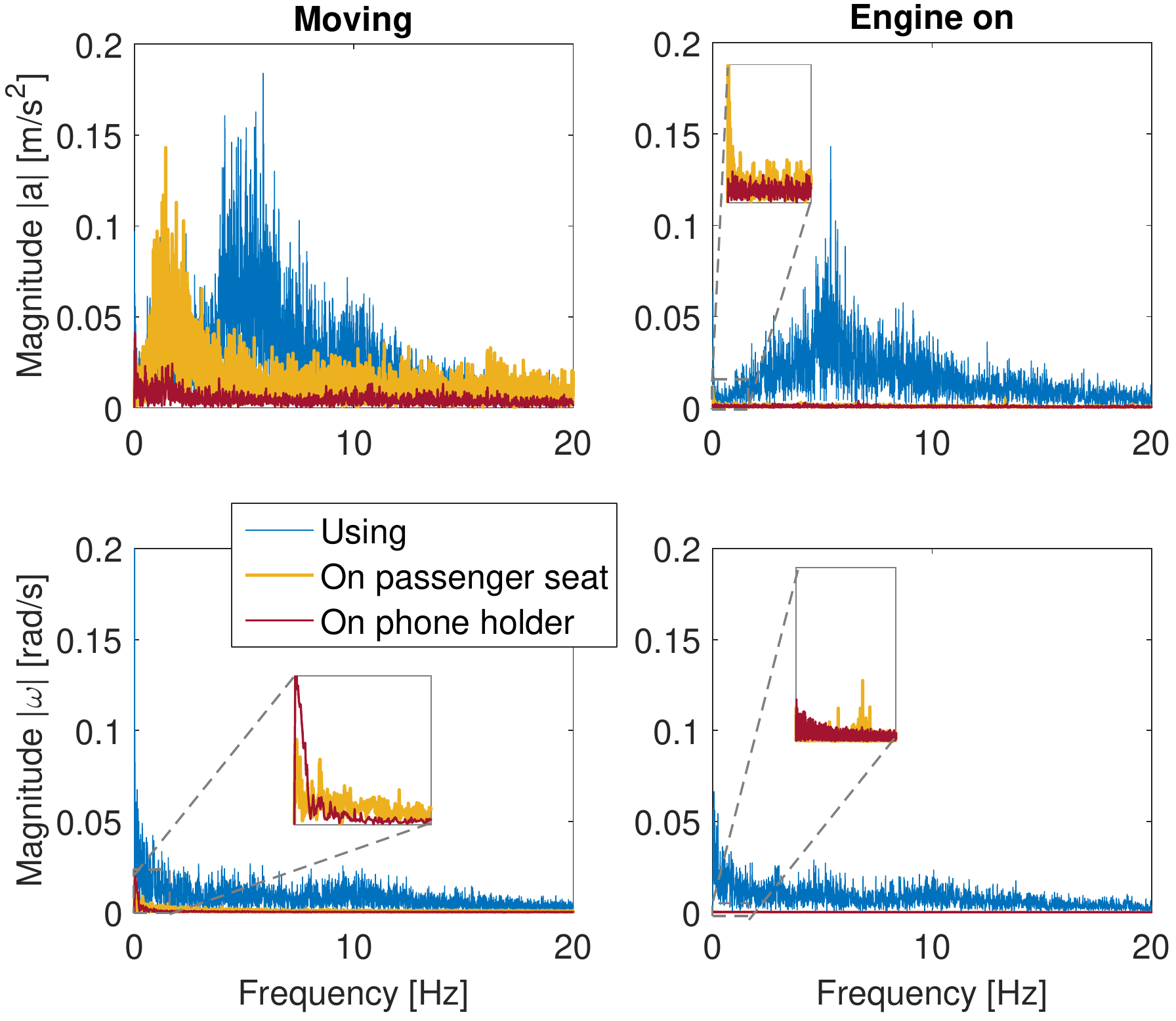}
 \caption{The spectra of the three use case scenarios have been computed when the vehicle is standing still (on the left) and when it is moving (on the right). In the top plot, the acceleration norm of the three scenarios is shown; in the bottom one, the angular velocity norm spectra are shown.}
 \label{fig:spectra_analysis}
\end{figure}

Although data intuitively exhibit some differences among the three scenarios, these discrepancies are not enough to discriminate the two classes (use and not use of the phone). Data need to be rearranged in order to become separable. To effectively manipulate them, the different scenarios are analyzed also in the frequency domain. In this context, differences are clearer, as depicted in Fig. \ref{fig:spectra_analysis}. In fact, analyzing the acceleration norm spectra, two main peaks are visible: one in the range  $1-3\ \mathrm{Hz}$, and a second one in the interval $4-12\ \mathrm{Hz}$. The first peak is related to the harmonics more influenced by the vehicle dynamics, not present when the vehicle is not moving. This condition is shared by both \textit{using} and \textit{on passenger seat} conditions, but it is less intense in the \textit{on phone holder} one. A possible explanation of this fact can be due to the filtering action performed by the holder mechanism, smoothing the movements induced by the vehicle motion. On the contrary, the second peak is present only in the \textit{using} case, no matter whether the vehicle is moving or standing still. These harmonics could thus be used as a signature to discriminate whether the driver is using the phone or not.

Instead of the noticeable differences in the spectrum of the acceleration norms, the gyroscope signals norm spectra do not show any prevalent peak. Nevertheless, it is apparent that the \textit{using} condition gives data that are more excited over all the frequencies (up to $20\ \mathrm{Hz}$), than in the other two scenarios, in which only few low-frequency harmonics differ from the baseline. This difference is, in addition to the one found for the acceleration norm, a characteristic signature of the \textit{in-use} case.

%%%%%%%%%%%%%%%%%%%%%%%%%%%%%%%%%%%%%%%%%%%%%%%%%%%%%%%%%%%%%%%%%%%%%%%%%%%%%%%%
%%%%%%%%%%%%%%%%%%%%%% Features extraction %%%%%%%%%%%%%%%%%%%%%%%%%%%%%%%%%%%%%
\subsection{Features extraction}
\begin{figure}[h]
\begin{center}
\subfloat[]{\includegraphics[width=0.45\textwidth]{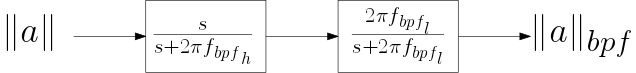}\label{fig:bpf_filter}}\\
\subfloat[]{\includegraphics[width=0.45\textwidth]{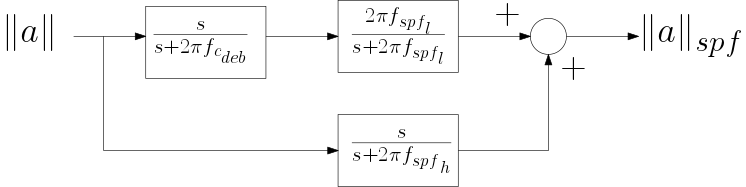}\label{fig:spf_filter}}\\ 
\subfloat[]{\includegraphics[width=0.45\textwidth]{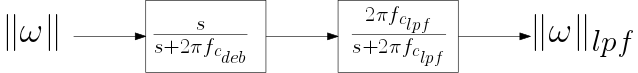}\label{fig:lpf_filter}}\\
\subfloat[]{\includegraphics[width=0.45\textwidth]{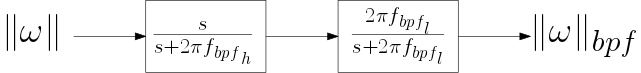}\label{fig:bpf_filter_gyro}}
\end{center}
\caption{Block diagrams of the four filters used to preprocess the acceleration and rotational speed norms. In Fig. \ref{fig:bpf_filter} and \ref{fig:bpf_filter_gyro}, the band-pass filters used to derive $\|a\|_{bpf}$ and $\|\omega\|_{bpf}$, respectively, are shown. Fig. \ref{fig:spf_filter} represents the filtering chain used to derive $\|a\|_{spf}$. Fig. \ref{fig:lpf_filter} depicts the low-pass filter used to get $\|\omega\|_{lpf}$.}
\label{fig:filters}
\end{figure}

The analysis in the frequency-domain provides insights on the main differences between \textit{in-use} and \textit{not in-use} classes. Features can be extracted with a data pre-processing phase, aimed at filtering the data in specific frequency-ranges in order to retain the most informative harmonics only. To do this, four new signals are defined:
\begin{itemize}
\item $\|a\|_{bpf}$ represents the acceleration norm filtered with a band-pass filter (Fig. \ref{fig:bpf_filter}) in the range $f_{bpf_{h}}=4\ \mathrm{Hz}\ -\ f_{bpf_{l}}=15\ \mathrm{Hz}$;
\item $\|a\|_{spf}$ represents the sum of the acceleration norm filtered with a low-pass filter (with cut-off frequency $f_{spf_{h}}=4\ \mathrm{Hz}$) and a high-pass filter ($f_{spf_{l}}=15\ \mathrm{Hz}$). To make this signal comparable to the previous one, its bias is removed with a high-pass filter with $f_{c_{deb}}=0.05 \ \mathrm{Hz}$ (Fig. \ref{fig:spf_filter});
\item $\|\omega\|_{lpf}$ represents the angular velocity norm filtered with a low-pass filter (Fig. \ref{fig:lpf_filter}), with cut-off frequency $f_{c_{lpf}}=20\ \mathrm{Hz}$. As for $\|a\|_{spf}$, the very-low frequency harmonics are removed with a high-pass filter with $f_{c_{deb}}=0.05 \ \mathrm{Hz}$;
\item $\|\omega\|_{bpf}$ represents the angular rate norm filtered with the same band-filtered used for $\|a\|_{bpf}$, removing the influence of the harmonics most affected by the vehicle dynamics (Fig. \ref{fig:bpf_filter_gyro}).
\end{itemize}

\begin{figure}[thpb]
 \centering
 \includegraphics[width=0.45\textwidth]{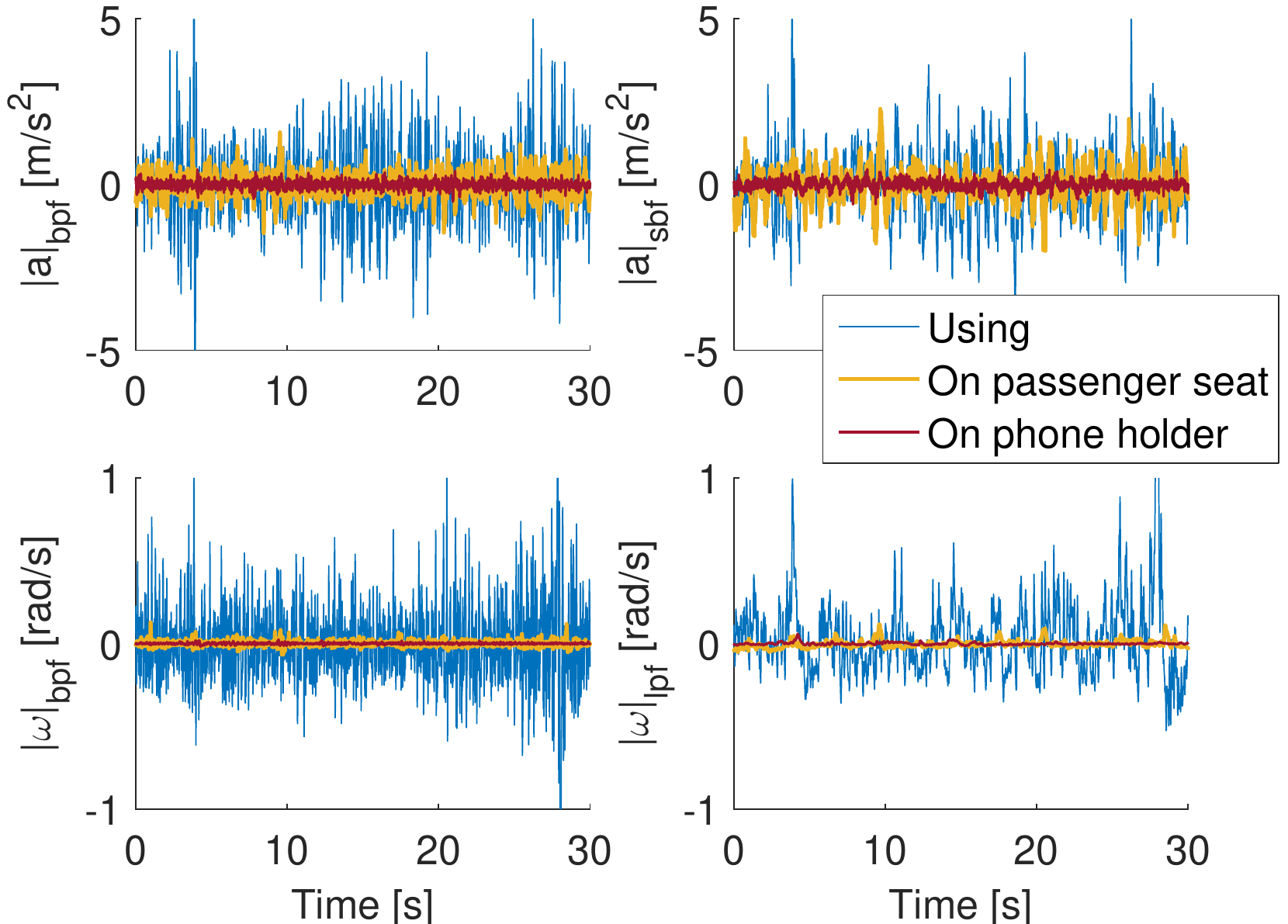}
 \caption{The acceleration and angular velocity norms: raw and filtered data. In all four situations, it is evident that the variance of the \textit{using} cases is larger than when the phone is not used.}
 \label{fig:features_filtered}
\end{figure}

\begin{figure}[thpb]
 \centering
 \includegraphics[width=0.45\textwidth]{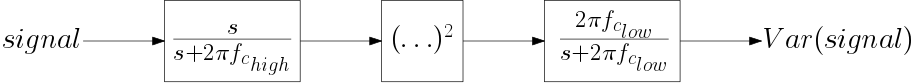}
 \caption{Estimation of the variance using a chain of LTI filters: the signal is first filtered with a high-pass filter with cut-off frequency $f_{c_{high}}$, removing the low frequency components linked to the mean value $signal_{mean}$; then, the filtered signal is squared, obtaining an estimate of $(signal-signal_{mean})^2$; finally, the expected value is obtained by filtering with a low-pass filter, getting the estimate of the variance $\EX\left((signal-signal_{mean})^2\right)$}
 \label{fig:variance_filters}
\end{figure}

The results of the pre-processing phase is shown in Fig. \ref{fig:features_filtered}. In time domain, the new four signals span different ranges, which are always smaller for the \textit{not in-use} cases than for \textit{in-use}. This difference can be quantified by estimating the variance of the filtered signals. This can be done online by computing the second order moment on a sliding window or, as illustrated in Fig. \ref{fig:variance_filters}, by means of a filtering chain: the signal is initially unbiased by removing the low-frequency harmonics with a high-pass filter ($f_{c_{high}}=0.01\ \mathrm{Hz}$), then squared, and finally the expected value is obtained with a low-pass filter ($f_{c_{low}}=0.1\ \mathrm{Hz}$).

\begin{figure}[thpb]
 \centering
 \includegraphics[width=0.45\textwidth]{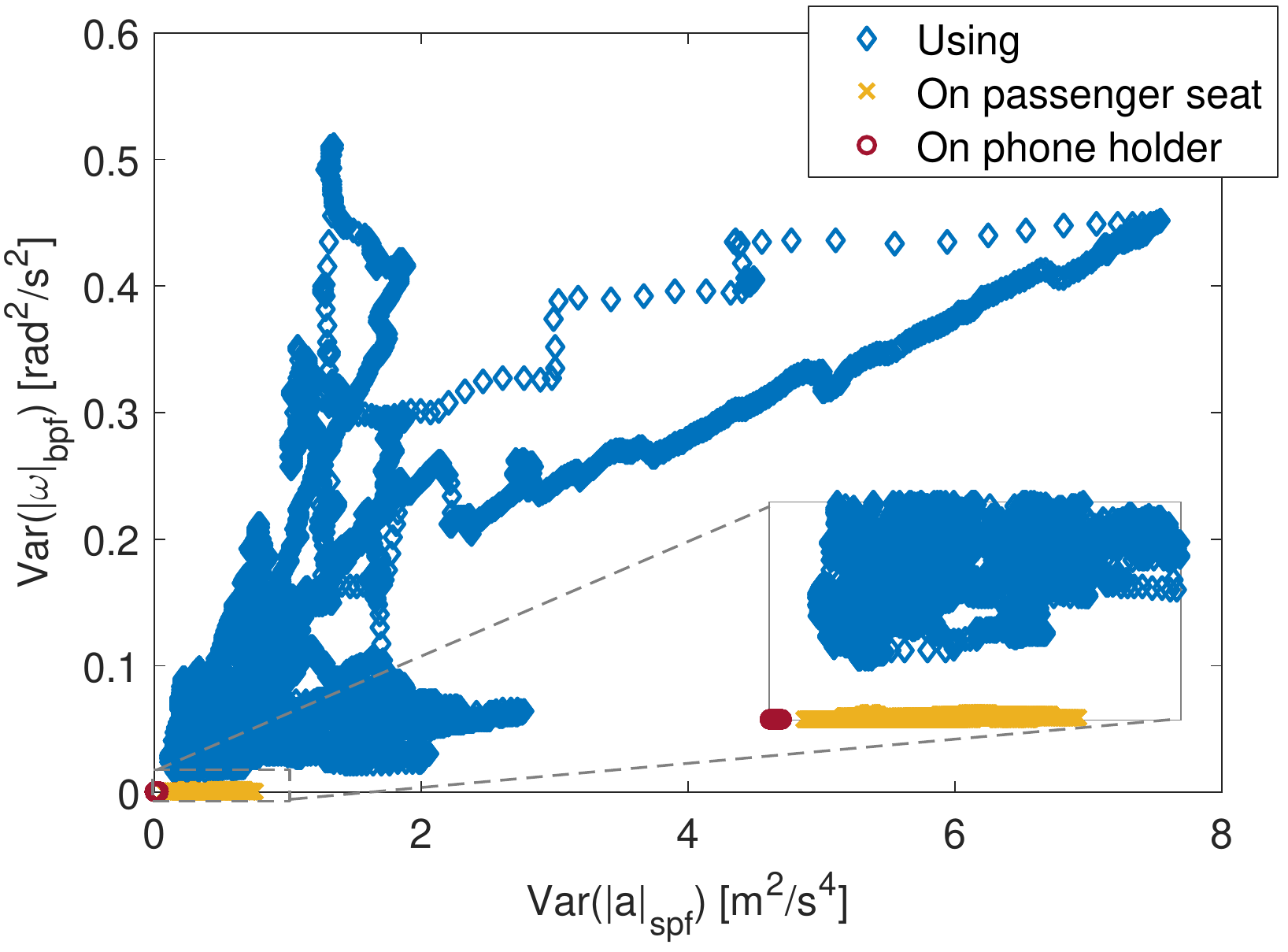}
 \caption{An example two selected features.}
 \label{fig:features_scattered}
\end{figure}

By estimating the variance of the four filtered signals, five features are extracted: $Var\left(\|a\|_{bpf} \right)$, $Var\left(\|a\|_{spf} \right)$, $Var\left(\|\omega\|_{bpf} \right)$, $Var\left(\|\omega\|_{lpf} \right)$, and $\EX \left(\frac{Var\left(\|a\|_{bpf} \right)}{Var\left(\|a\|_{spf} \right)} \right)$, the mean value of the ratio between the variance of $\|a\|_{bpf}$ and $\|a\|_{spf}$. As illustrated in the example of Fig. \ref{fig:features_scattered}, the \textit{in-use} and \textit{not in-use} classes are finally separated. However, the most informative among the proposed ones are still to be singled out. Since the number of potential features is relatively small, the optimal ones are selected following a so-called \textit{filter approach} \cite{liu2007computational}, \textit{i.e.}, iterating a training and performance evaluation phase and maximizing the classification performance.

%%%%%%%%%%%%%%%%%%%%%%%%%%%%%%%%%%%%%%%%%%%%%%%%%%%%%%%%%%%%%%%%%%%%%%%%%%%%%%%%
%%%%%%%%%%%%%%%%%%%%%% Classification algorithm %%%%%%%%%%%%%%%%%%%%%%%%%%%%%%%%%%%%%
\section{Classification algorithm} \label{algorithm}
Once the features are extracted, the classification problem can be defined. In this work, as shown in the example of the previous section, features have proved to be linearly separable. Thus, to optimize the class separation, the SVM algorithm is used. SVM is preferred, among all the linear classifiers, for its relatively small computational effort during the prediction, a crucial phase for a reduced computational power and limited energy storage device, such as the target smartphone platform.

SVM is a widely known kernel-based Machine Learning (ML) algorithm, which finds the optimal separating hyperplane by maximizing the separation margin between any training point and the hyperplane itself  \cite{scholkopf2001learning,duda2012pattern,friedman2001elements}. The optimal hyperplane is obtained solving
\begin{equation}
\begin{split}
\min_{W\in \mathcal{H},\ b\in \mathbb{R},\ \xi\in \mathbb{R}}\ & \frac{1}{2} \|W\|^2+C\sum_{i=1}^{m}\xi_i\\
subject\ to\ & y_i\left(\langle w,x_i \rangle +b \right)\ge1-\xi_i\ \forall i=1,\dots,m\\
&\xi_i\ge0
\end{split}
\label{SVM}
\end{equation}
where $W$ is the vector orthogonal to the hyperplane, $\mathcal{H}$ is the product space, $(x_i,\ y_i)$ are the training inputs and outputs, and $C>0$ is the penalty parameter of the error term $\xi$, which avoids the classification to be biased by outliers.

%%%%%%%%%%%%%%%%%%%%%%%%%%%%%%%%%%%%%%%%%%%%%%%%%%%%%%%%%%%%%%%%%%%%%%%%%%%%%%%%
%%%%%%%%%%%%%%%%%%%%%% Results %%%%%%%%%%%%%%%%%%%%%%%%%%%%%%%%%%%%%
\section{Results} \label{results}
\begin{figure}[thpb]
 \centering
 \includegraphics[width=0.45\textwidth]{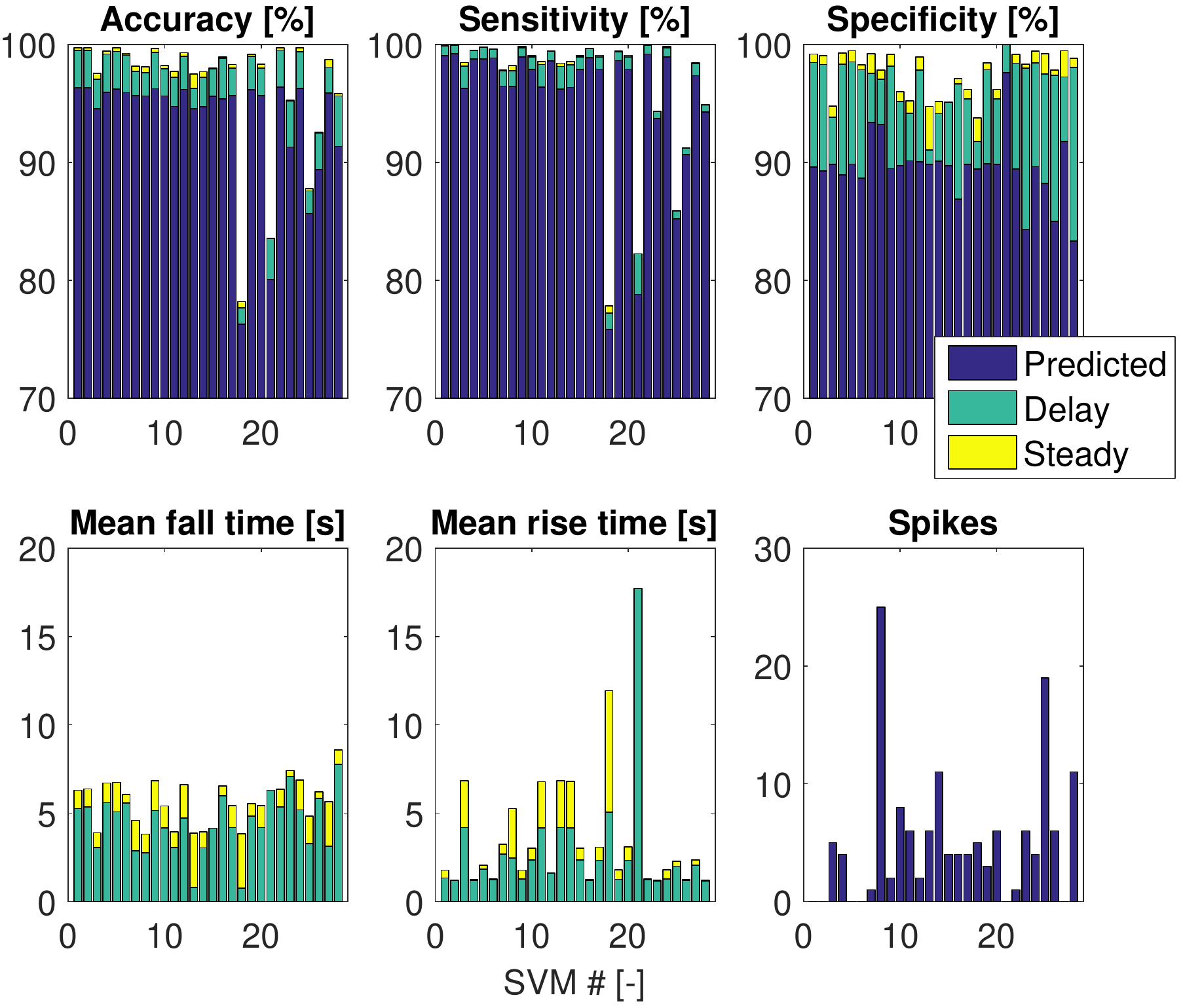}
 \caption{Performance of the trained SVM against the validation dataset.}
 \label{fig:validation}
\end{figure}

The SVM algorithm is trained with the all the features illustrated in Section \ref{features_selection}. The training dataset is composed of several hour-long tests recorded during mixed driving conditions (\textit{i.e.}, mixed urban and highway), in which the phone has been used reproducing the daily user experience. Among all the 31 possible features combinations, only 28 provide satisfactory results. In Fig. \ref{fig:validation}, the performance in validation is analyzed in terms of accuracy, specificity and sensitivity. The SVM 22, trained only with $Var(\|\omega\|_{bpf})$, shows the highest accuracy ($96.38\%$), with the second highest sensitivity ($99.19\%$) and an acceptable specificity ($89.44\%$).

\begin{figure}[thpb]
 \centering
 \includegraphics[width=0.45\textwidth]{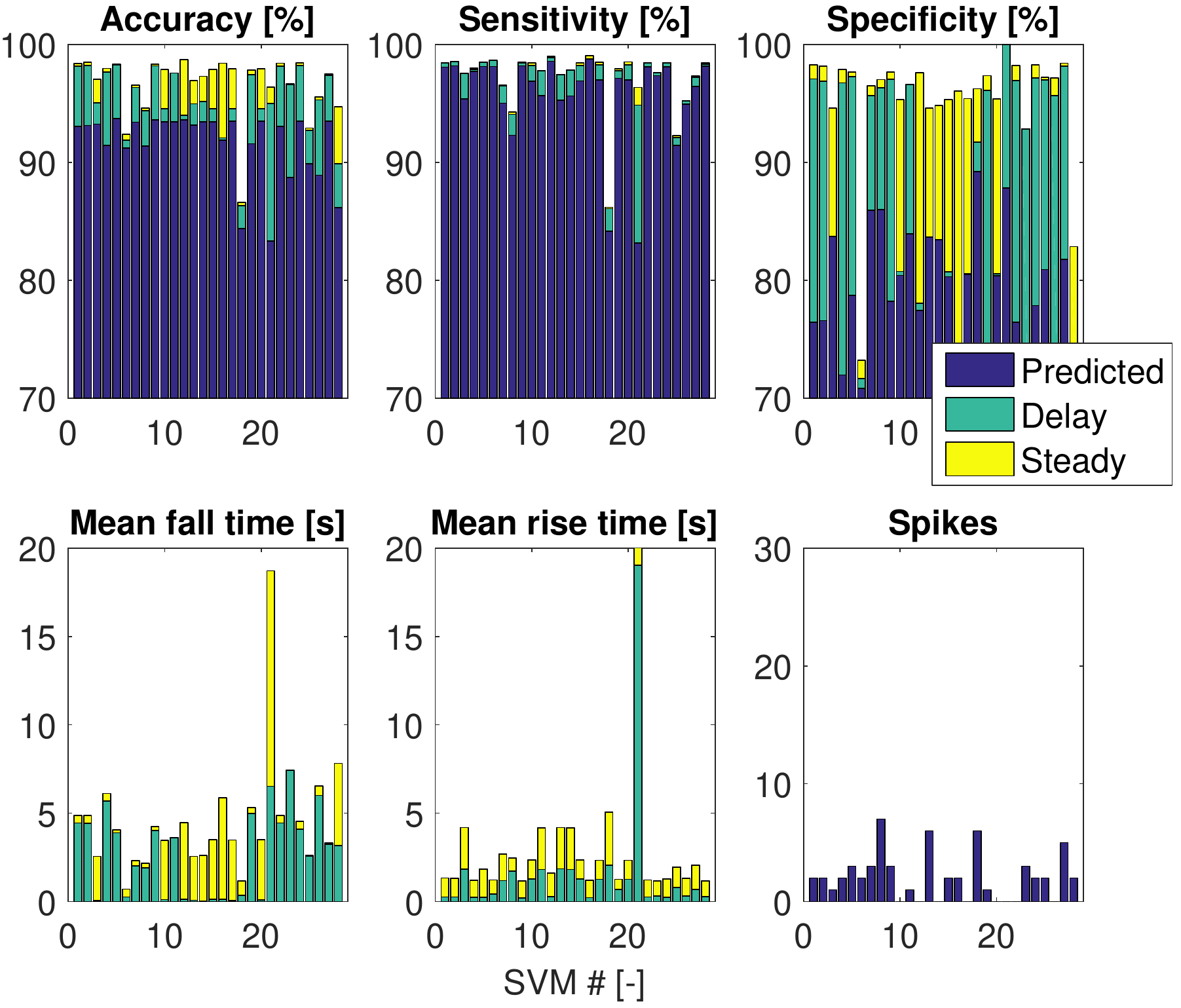}
 \caption{Performance of the trained SVM against a different dataset, used for testing.}
 \label{fig:testing}
\end{figure}

For a fair evaluation of the performance, the same SVMs are tested against a different dataset (Fig. \ref{fig:testing}). Although performance slightly drop for all the trained models, SVM 22 - the most performing classifier in validation - is still the best, achieving $93.053\%$ in accuracy, $98.116 \%$ in sensitivity, with a drop in specificity ($76.431 \%$) though.

One of the strongest assumptions behind the vast majority of ML algorithms is that data are independent. It is obviously not true in this application, as correlations are inherently introduced by the motion constraints and because of the procedure with which features are extracted (\textit{i.e.}, variance is, in a nutshell, evaluated over a moving window, which intrinsically introduces correlation). The limits of the classifier become more evident during transients from one class to another: classification is delayed by the effect of such transients and, therefore, the classification performance degrades. 

%The limits of the classifier become more evident during transients from one class to another, as shown in Fig. \ref{fig:validation_time}. The classification is delayed by the effect of such transients and, therefore, the classification performance degrades. 

To better understand whether the classifier is well performing, the duration of the transients is evaluated: \textit{delay} represents the mean time between the label variation and the corresponding first change in the classification; instead, \textit{steady} is the mean time to reach a steady-state output. As shown in Fig. \ref{fig:validation}, the classifiers are generally faster in the detection phase when the user starts using the phone (denoted as \textit{rise time}) than when she/he stops using it (\textit{fall time}).

Having computed average time of the transients, classification performance can be evaluated removing their effects of on the prediction. Results improve significantly: in validation, SVM 22 obtains an accuracy of $99.74\%$, the highest sensitivity of $99.95\%$ and an important increase of the specificity, which is now $99.17\%$; against testing data, the accuracy reaches $98.427 \%$, sensitivity $98.47\%$, and specificity $98.213\%$, a consistent improvement with respect to the nominal condition. 

Another, though less relevant, side effect of testing the SVM on dynamics data is the presence of spikes. Spikes can be defined as very short-time outliers in the prediction, too short to be realistic according to the underlying dynamics (\textit{e.g.}, in this application, spikes are not longer than few samples, a very unlikely use of the phone). As spikes are very short, they do not play a significant role in the evaluation of the main classification indexes (accuracy, specificity, sensitivity), but it is still interesting to compare the number of their occurrences for the different classifiers.

\begin{table*}[htbp]
\begin{center}
\scalebox{0.85}{
\begin{tabular}{|c|c|c|c|c|c|c|c|c|c|c|}
\hline
\multirow{2}{*}{\textbf{Dataset}}&\multirow{2}{*}{\textbf{Classifier}}&\textbf{Features}&\multicolumn{2}{|c|}{\textbf{Accuracy}}&\multicolumn{2}{|c|}{\textbf{Sensitivity}}&\multicolumn{2}{|c|}{\textbf{Specificity}}&\multirow{2}{*}{\textbf{Time rise steady [s]}}&\multirow{2}{*}{\textbf{Time fall steady [s]}}\\
\cline{4-9}
&&\textbf{Used}&Nominal&Steady&Nominal&Steady&Nominal&Steady&&\\
\hline
\multirow{10}{*}{Validation}&22& $\|\omega \|_{bpf}$&96.38&99.74&99.19&99.95&89.44&99.17&1.278&6.364\\
\cline{2-11}
&\multirow{2}{*}{24}& $\|\omega \|_{bpf}$&\multirow{2}{*}{96.28}&\multirow{2}{*}{99.73}&\multirow{2}{*}{98.936}&\multirow{2}{*}{99.216}&\multirow{2}{*}{89.63}&\multirow{2}{*}{99.449}&\multirow{2}{*}{1.781}&\multirow{2}{*}{6.882}\\
&& $\|\omega \|_{lpf}$&&&&&&&&\\
\cline{2-11}
&\multirow{3}{*}{5}& $\|\omega \|_{bpf}$&\multirow{3}{*}{96.243}&\multirow{3}{*}{99.72}&\multirow{3}{*}{98.769}&\multirow{3}{*}{99.815}&\multirow{3}{*}{89.851}&\multirow{3}{*}{99.472}&\multirow{3}{*}{2.042}&\multirow{3}{*}{6.7391}\\
&& $\|a \|_{spf}$&&&&&&&&\\
&& $\|\omega \|_{lpf}$&&&&&&&&\\
\cline{2-11}
&\multirow{2}{*}{1}& $\|\omega \|_{bpf}$&\multirow{2}{*}{96.346}&\multirow{2}{*}{99.71}&\multirow{2}{*}{99.055}&\multirow{2}{*}{99.9}&\multirow{2}{*}{89.599}&\multirow{2}{*}{99.203}&\multirow{2}{*}{1.771}&\multirow{2}{*}{6.294}\\
&& $\|a \|_{spf}$&&&&&&&&\\
\cline{2-11}
&\multirow{2}{*}{2}& $\|\omega \|_{bpf}$&\multirow{2}{*}{96.351}&\multirow{2}{*}{99.71}&\multirow{2}{*}{99.223}&\multirow{2}{*}{99.95}&\multirow{2}{*}{89.28}&\multirow{2}{*}{99.056}&\multirow{2}{*}{1.209}&\multirow{2}{*}{6.375}\\
&& $\|a \|_{bpf}$&&&&&&&&\\
\hline

\multirow{10}{*}{Testing}&22& $\|\omega \|_{bpf}$&93.053&98.421&98.116&98.47&76.431&98.213&0.2518&4.877\\
\cline{2-11}
&\multirow{2}{*}{24}& $\|\omega \|_{bpf}$&\multirow{2}{*}{93.479}&\multirow{2}{*}{98.427}&\multirow{2}{*}{98.126}&\multirow{2}{*}{98.46}&\multirow{2}{*}{77.852}&\multirow{2}{*}{98.288}&\multirow{2}{*}{0.23}&\multirow{2}{*}{4.537}\\
&& $\|\omega \|_{lpf}$&&&&&&&&\\
\cline{2-11}
&\multirow{3}{*}{5}& $\|\omega \|_{bpf}$&\multirow{3}{*}{93.741}&\multirow{3}{*}{98.326}&\multirow{3}{*}{98.144}&\multirow{3}{*}{98.483}&\multirow{3}{*}{78.727}&\multirow{3}{*}{97.676}&\multirow{3}{*}{0.232}&\multirow{3}{*}{4.059}\\
&& $\|a \|_{spf}$&&&&&&&&\\
&& $\|\omega \|_{lpf}$&&&&&&&&\\
\cline{2-11}
&\multirow{2}{*}{1}& $\|\omega \|_{bpf}$&\multirow{2}{*}{93.041}&\multirow{2}{*}{98.407}&\multirow{2}{*}{98.087}&\multirow{2}{*}{98.442}&\multirow{2}{*}{76.442}&\multirow{2}{*}{98.26}&\multirow{2}{*}{0.252}&\multirow{2}{*}{4.86}\\
&& $\|a \|_{spf}$&&&&&&&&\\
\cline{2-11}
&\multirow{2}{*}{2}& $\|\omega \|_{bpf}$&\multirow{2}{*}{93.142}&\multirow{2}{*}{98.493}&\multirow{2}{*}{98.202}&\multirow{2}{*}{98.568}&\multirow{2}{*}{76.569}&\multirow{2}{*}{98.183}&\multirow{2}{*}{0.254}&\multirow{2}{*}{4.869}\\
&& $\|a \|_{bpf}$&&&&&&&&\\
\hline
\end{tabular}}
\end{center}
\caption{Comparison of the performance of the most performing classifiers against validation and testing data.}
\label{tab:classifiers}
\end{table*}

%\begin{figure}[thpb]
% \centering
% \includegraphics[width=0.45\textwidth]{validation_testing_time.pdf}
 %\caption{An overview of the algorithm performance validation over time.}
 %\label{fig:validation_time}
%\end{figure}

To better analyze the achieved classification performance, the five most performing classifiers are compared in Table \ref{tab:classifiers}:
\begin{itemize}
\item the performance of these five classifiers is comparable, both in the nominal and steady-state cases, with limited differences also for the mean rise and fall time;
\item the five most performing classifiers have $\|\omega \|_{bpf}$ as main feature. This means that gyroscopes play a key-role in the detection. The most performing SVM which uses only accelerometers (SVM 8 - trained with $\|a\|_{bpf}$) achieves a nominal accuracy of $95.64 \%$ (which is $98.12 \%$ at steady-state). Although still very performing, smartphones not equipped with gyroscopes could not achieve the best possible performance;
\item the short rise time means that the classifier is very prompt in detecting when the driver starts using the phone. However, the classifier is slower in detecting when the phone is no longer used. As the detection of the use of the phone is pursued only to profile driver's behavior, these few seconds of use do not significantly alter the final risk index (especially in long journeys).
\end{itemize}

\section{Concluding remarks} \label{conclusions}
This paper showed that, using only sensors commonly installed on a smartphone, the device itself can be employed to monitor whether it is used when driving. This is of particular importance to assess risky behaviors of drivers, as phone usage is one of the main sources of accidents and fatalities. The proposed approach relies on a data-processing and SVM-based classification, and proved to be effective under realistic use. Crucial is extracting features in the frequency-domain, separating the influence of vehicle dynamics from the phone using one. Future work will tackle the same problem using a more data-driven approach, reducing the effort request tuning the digital signal process phase, by means of more sophisticated models (e.g., deep-learning) or investigating innovative automatic features extraction methods.

%%%%%%%%%%%%%%%%%%%%%%%%%%%%%%%%%%%%%%%%%%%%%%%%%%%%%%%%%%%%%%%%%%%%%%%%%%%%%%%%
%%%%%%%%%%%%%%%%%%%%%%%%%%% Acknowledgment %%%%%%%%%%%%%%%%%%%%%%%%%%%%%%%%%%%%%
\section*{Acknowledgments}
The authors gratefully acknowledge the help of Fabio Martellotta with the experimental data collection activities.

\IEEEpeerreviewmaketitle
\bibliographystyle{ieeetr}
\bibliography{Kirey_smartphone_use_detection_v2}

\end{document}